\documentclass{article}
\usepackage{graphicx}  
\usepackage{amsmath}   
\usepackage[compress]{cite}
\usepackage{amssymb}   
\usepackage{bm} 
\usepackage{dcolumn}
\usepackage{color}
\usepackage{mathrsfs}
\usepackage{amsfonts}
\usepackage{varioref}
\usepackage{color}
\RequirePackage[colorlinks,citecolor=blue,urlcolor=magenta,linkcolor=blue]{hyperref}
\labelformat{section}{Section #1} 
\labelformat{subsection}{Section #1} 
\labelformat{subsubsection}{Section #1}
\labelformat{subsubsubsection}{Section #1}
\labelformat{equation}{Eq.~(#1)} \labelformat{figure}{Fig.~#1} 
\labelformat{subfigure}{Fig.~\thefigure#1} 
\labelformat{table}{Tab.~#1} 
\labelformat{appendix}{Appendix #1}
\title{\bf Black Hole Topology in $f(R)$ Gravity}
\author{Akash K Mishra \footnote{akash.mishra@iitgn.ac.in}$~^{1}$, Mostafizur Rahman \footnote{mostafizur@ctp-jamia.res.in}$~^{2}$ and Sudipta Sarkar\footnote{sudiptas@iitgn.ac.in}$~^{1}$\\
\small{$~^{1}$ Indian Institute of Technology, Gandhinagar-382355, Gujarat, India}\\
\small{$~^{2}$ Centre for Theoretical Physics, Jamia Millia Islamia, New Delhi-110025, India}\\
}

\begin{document}
  
\maketitle
\begin{abstract}

Hawking's topology theorem in general relativity restricts the cross-section of the event horizon of a black hole in $3+1$ dimension to be either spherical or toroidal. The toroidal case is ruled out by the topology censorship theorems. In this article, we discuss the generalization of this result to black holes in $f(R)$ gravity in $3+1$ and higher dimensions. We obtain a sufficient differential condition on the function $f'(R)$, which restricts the topology of the horizon cross-section of a black hole in $f(R)$ gravity in $3+1$ dimension to be either $S^2$ or $S^1\times S^1$. We also extend the result to higher dimensional black holes and show that the same sufficient condition also restricts the sign of the Yamabe invariant of the horizon cross-section.

\end{abstract}
\section*{Introduction}

Black holes have often provided profound insights into the nature of quantum gravity and the structure of space-time. One of the most intriguing results in classical black hole physics is the topology theorem by Hawking \cite{1,2}, which asserts that the cross-section of the event horizon in a stationary, asymptotically flat space-time in four-dimensions, with matter stress-energy tensor obeying dominant energy condition, is always topologically $S^2$ or $S^1 \times S^1$. The proof beautifully relates the Euler characteristics $\chi$, associated with the horizon cross-section and rate of change of expansion of a congruence of outgoing future directed null geodesic. It shows if the topology is not spherical or toroidal, then one can always have a trapped surface outside the event horizon, which is a contradiction.  \\

This result, however, doesn't trivially generalize either to the higher dimensions or other modified theories of gravity. This is because in a four-dimensional space-time the cross-section of the event horizon is a co-dimension two surface and the Euler characteristics is related to the intrinsic curvature through Gauss-Bonnet theorem. While in higher dimensions it is not the case since the Gauss-Bonnet theorem can't be applied in the same form. But instead of Euler number, one can associate a Yamabe number to any co-dimension two surface in any dimensions. Work by Galloway and Schoen \cite{Galloway:2005mf, Galloway:2011np} shows that there is still a restriction on the sign of the Yamabe invariant and as a result, even in higher dimensions, only certain classes of horizon topology are allowed. Such kind of topological restrictions on a higher dimensional black hole is extensively reviewed by Emparan and Reall \cite{Emparan:2008eg}, which shows how the topology of horizon cross-section of a black hole in a five-dimensional stationary vacuum space-time to be either $S^3$ or $S^2\times S^1$ (known as a Black Ring solution). Hence it is quite evident that higher dimensional black holes are more elegant and richer regarding topological structures as compared to the lower dimensional ones.  \\

The restriction on the topology of the event horizon is ultimately linked with the topology censorship theorems which constrains the topology of the space-time outside the event horizon. In general relativity, it is possible to show that the topology of the event horizon can be obtained from the censorship theorems\cite{Friedman:1993ty} even without the stationarity\cite{Jacobson:1994hs}. This result also asserts that general relativity does not allow an observer to probe the topology of space-time, all non-trivial topological structures are hidden inside the horizon. \\

The topology of black holes for modified gravity theories such as a  $f(R)$ theory, is still unexplored. The equation of motion has changed, and the imposition of dominant energy condition may not be enough to restrict the topology. For $f(R)$  gravity, a simpler approach could be to transform the theory into general relativity with extra matter fields using a conformal transformation. Under such a conformal transformation, a stationary event horizon mapped to itself \cite{Jacobson:1993pf} and the topology remains invariant. Then, it could be possible to understand the topology of stationary black holes in $f(R)$ gravity from the corresponding result in general relativity.\\

The understanding of the topology of black hole horizons beyond general relativity could be an important tool to distinguish a modified gravity theory from general relativity. We expect that the space of solutions of any modified gravity theory would be richer and there could be new solutions with no parallel in general relativity. The simplest of such a modified theory is the $f(R)$ gravity, where we add higher curvature corrections which are functions of Ricci scalar only to the Einstein Hilbert action. In this work, we investigate the topology of stationary black holes in $f(R)$ theory in both $3+1$ and higher dimensions. In $3+1$ case, we find a sufficient differential condition (along with dominant energy condition) on the form of the function $f(R)$ that makes the stationary black hole topology same as that of general relativity. By following the procedure as in \cite{Galloway:2005mf, Galloway:2011np}, we show that the same condition also restricts the sign of Yamabe invariant of higher dimensional stationary black holes.  This result is interesting because it provides a sufficient condition on the choice of $f(R)$ theories for which the topology of the black hole event horizon is same as general relativity, irrespective of the dimension of space-time. We also provide an alternative derivation of the same condition using the conformal transformation technique.

 \section*{Topology of a stationary black hole horizon in $3+1$ dimensions}

We consider first an event horizon in $3+1$ dimensions in a stationary  space-time with metric $g_{\mu \nu}$. The event horizon $\mathcal{H}$ in a stationary space-time is also a Killing horizon\cite{2} on which the norm of a Killing vector (which is time-like outside) vanishes. We may consider null Killing field as the generator $k^\mu$ of the event horizon which obeys the non-affine geodesic equation: $k^\mu \nabla_\mu k^\nu = \kappa \, k^\nu$, where $\kappa$ represents the surface gravity. The cross-section of the event horizon is a compact space-like surface $\mathcal{B}$. If $n^\mu$ is the auxiliary null normal to the horizon then the intrinsic metric of the cross-section is: $\beta_{\mu\nu} = g_{\mu\nu} + k_\mu n_\nu + k_\nu n_\mu$,  where we have used the normalization condition $ k_\mu n^\mu = -1$. The Euler characteristics of such a surface is then given by,

\begin{equation}\label{Eulerch}
\chi(\mathcal{B}) = \int_{\mathcal{B}} \mathcal{K}_G\,  d\mathcal{A}
\end{equation}
\\ 
Where $\mathcal{K}_G$ is the Gaussian curvature of the cross-section $\mathcal{B}$ of the horizon $\mathcal{H}$. For a stationary space-time, the horizon has vanishing expansion and shear. Then, the intrinsic Gaussian curvature, can be related to the full space-time curvature \cite{Gourgoulhon:2005ng,9} and the Euler characteristics is expressed as,
 
\begin{equation}
\chi(\mathcal{B})=\int_{\mathcal{B}} \left(  R + 4\,R_{\mu \nu} k^\mu n^\nu - 2\,R_{\mu \nu \alpha \beta} k^\mu n^\nu k^\alpha n^\beta \right) d\mathcal{A} \label{chi_def}
\end{equation}
 \\
This equation relates the topological information of the horizon to the full space-time geometry. Since in any theory of gravity, given the matter content, the field equation determines the space-time geometry, we expect to find a constraint on the topology of the horizon from the constraint on the matter stress tensor. \\

Hawking's Topology theorem\cite{1} assumes that the stress-energy tensor to satisfy the dominant energy condition and the space-time metric to be a solution of Einstein's equation. Then, the Euler characteristics is shown to be always positive semi definite. For a two dimensional compact orientable surface, this leads to only two possibilities. The cross-section of the horizon could be either a sphere $S^2$ or a torus $ S^1 \times S^1$. The case for torus topology turns out to be unstable\cite{Jacobson:1994hs,Friedman:1993ty} and the only physical possibility is the $2$-sphere. \\

In this work, we are interested in the topological structure of horizon cross-section arising from higher curvature theories of the form,
\begin{equation}
I_0 = \int d^4x \sqrt{-g} \left[\frac{f(R)}{16\pi} +\mathcal{L}_m \right]
\end{equation}
\\
Here $\mathcal{L}_m$ represents the Lagrangian for matter fields and $f(R)$ is some function of Ricci scalar. Variation of this Action with respect to metric yields the field equation~\cite{Sotiriou:2008rp,Guarnizo:2010xr},

 \begin{equation}
f'(R)R_{\mu\nu}-\frac{1}{2}g_{\mu\nu}f(R)+g_{\mu\nu}\square f'(R) -\triangledown_\mu \triangledown_\nu f'(R) = 8\pi T_{\mu\nu}
\end{equation}

$f'(R)$ denotes the derivative of the function $f(R)$ with respect to its argument. Let us assume that the matter energy tensor to obey the Dominant energy condition. Also using \ref{chi_def} and the field equation of $f(R)$ gravity, the Euler characteristics of the horizon cross-section for such a theory has expression of the following form,

\begin{eqnarray} \label{chif}
\chi(\mathcal{H}) &=& \int_{\mathcal{B}} \left(2R_{\mu \nu} k^\mu n^\nu - 2R_{\mu \nu \alpha \beta} k^\mu n^\nu k^\alpha n^\beta \right) dA \nonumber \\
 &+& \int_{\mathcal{B}} \left[ \frac{R f'(R)-f(R)}{f'(R)} + \frac{2 k^\mu n^\nu \triangledown_\mu \triangledown_\nu f'(R) + 2\, \square f'(R)}{f'(R)}  + \left(\frac{16\pi \,T_{\mu \nu} k^\mu n^\nu}{f'(R)}\right)\right]dA \label{Euler_ch}
\end{eqnarray}

The general relativity limit can easily be obtained by setting $f(R) = R$. We first like to study the first integral, which is independent of the structure of the field equation. To tackle this term, we consider the variation of the null expansion $\theta = \beta^\nu_\mu \nabla_\nu k^\mu$, of an outgoing null congruence generated by null vector $k^\mu$ along the ingoing null geodesic generated by $n^\mu$ (parametrized by $s$). A straightforward calculation shows \cite{9},

\begin{equation}
R_{\mu \nu} k^\mu n^\nu - R_{\mu \nu \alpha \beta} k^\mu n^\nu k^\alpha n^\beta = -\frac{d\theta}{ds}-\beta^\mu_\sigma \beta^\nu_\rho \nabla_\mu n^\rho \nabla_\nu k^\sigma +p_\mu p^\mu - d^\dagger p\label{first_term}
\end{equation}

where $p^\mu =- n^\sigma \beta^{\mu \nu} \nabla_\nu k_\sigma$ and $d^\dagger p = -\beta^{\mu\nu}\nabla_\mu p_\nu$. Also, $p_\mu p^\mu$ is positive definite, since $p^\mu$ is a space-like vector field. The second term on the right hand side of the above equation represents shear and expansion on the horizon and vanishes by the stationary assumption. Integrating \ref{first_term} over the cross-section $\mathcal{B}$ gives,

$$\int_\mathcal{B} [R_{\mu \nu} k^\mu n^\nu -  R_{\mu \nu \alpha \beta} k^\mu n^\nu k^\alpha n^\beta] dA = \int_\mathcal{B} \left [p_\mu p^\mu-\frac{d\theta}{ds}\right] dA$$

Hence, in terms of the variation of null expansion along $n$ and the newly defined vector field $p$, the Euler characteristics of $\mathcal{B}$, takes the form,

\begin{align}
\chi(\mathcal{H}) &= \int_\mathcal{B} 2\left[p_\mu p^\mu-\frac{d\theta}{ds}\right]  dA\nonumber \\
& +\int_\mathcal{B} \left[ \frac{R f'(R)-f(R)}{f'(R)} + \frac{2 k^\mu n^\nu \triangledown_\mu \triangledown_\nu f'(R) + 2\square f'(R)}{f'(R)}  + \left(\frac{16\pi T_{\mu \nu} k^\mu n^\nu}{f'(R)}\right)\right]dA \label{eu_ch}
\end{align}

Since the event horizon of a stationary black hole is also the limit to the existence of outer trapped surfaces, when we approach the horizon from the domain of outer communication, the expansion coefficient $\theta$ takes positive value outside and vanishes on the horizon. This fixes the sign of  $\frac{d\theta}{ds}$ to be negative on $\mathcal{B}$ and as a result the first integral of the above equation turns out to be positive definite\cite{1,9} irrespective of the structure of field equation. \\
 
Now we turn our attention to the second integral of \ref{eu_ch}, which contains terms that are explicitly dependent on the form of field equation and apart from $T_{\mu \nu} k^\mu n^\nu$, one can't in general comment on the sign of any other terms. 
However, by doing a field redefinition as follows, one can identify some of the terms to be positive.  This is done by rewriting the action $I_0$ in a slightly different form, where an auxiliary scalar field $\phi$ is coupled to Ricci scalar as,
 
$$I_1 = \int d^4x \sqrt{-g}\left[\frac{f(\phi) +(R-\phi)f'(\phi)}{16\pi}+\mathcal{L}_m \right] $$

This leads to the equation of motion of the scalar field to be $\phi=R$ and one can use this to recover the initial action, i.e. $I_1(\phi=R) = I_0$. The dynamics of field equation obtained from $I_0$ is equivalent to that of $I_1$ ~\cite{Wands:1993uu}.
$I_1$ can now be easily transformed into Einstein gravity coupled to a scalar field by a conformal transformation, 

\begin{equation}
\bar{g}_{\mu\nu} = f'(\phi) g_{\mu\nu}\label{con_tr}
\end{equation}

which gives an action of the form~\cite{Jacobson:1995uq},

$$I_2 = \int d^4 x \sqrt{-\bar{g}} \frac{1}{16\pi}\left[\bar{R} -\frac{3}{2}\left(\frac{f''(\phi)}{f'(\phi)}\right)^2 \bar{\nabla}_\mu\phi \bar{\nabla}^\mu\phi + \frac{1}{f'(\phi)} \lbrace [f(\phi)-\phi f'(\phi)] + 16\pi L_m \rbrace \right]\label{I2}$$

The field Equation arising from action $I_2$ is,
\begin{equation}
\bar{R}_{\mu\nu}-\frac{1}{2} \bar{g}_{\mu\nu}\bar{R} = \frac{8\pi}{f'(\phi)}T_{\mu\nu} + \frac{3}{2}\left[\frac{f''(\phi)}{f'(\phi)} \right]^2 \bar{\nabla}_\mu\phi \bar{\nabla}_\nu\phi -\frac{3}{4} \bar{g}_{\mu\nu}\left[\frac{f''(\phi)}{f'(\phi)} \right]^2  \bar{\nabla}_\alpha\phi \bar{\nabla}^\alpha\phi +\frac{1}{2}\bar{g}_{\mu\nu} \frac{f(\phi) - \phi f'(\phi)}{f'(\phi)^2}\label{conformal}
\end{equation}

Considering the entire expression on the right hand side of \ref{conformal} as an effective energy-momentum tensor, it can be shown to satisfy the dominant energy condition, provided that, $f'(\phi)>0$ and $\phi f'(\phi)-f(\phi) \geq0$. Since the equation for the auxiliary scalar field $\phi$ is just, $\phi=R$, the above two conditions translates to $f'(R)>0$ and $Rf'(R)-f(R) \geq0$ in the physical frame~\cite{Jacobson:1995uq}. \\

The first inequality must hold, in order to have smooth conformal mapping between the higher curvature theory and the Einstein plus scalar theory. Also, in any $f(R)$ theory of gravity, $f'(R)$, represents the local entropy density of a black hole solution\cite{Jacobson:1995uq} and hence has to be positive definite. If this condition is violated, i.e for any process, the black hole space-time, starting with a configuration $f'(R)>0$, evolves to $f'(R)<0$, then $f'(R)$ has to be zero at some point in between and for that case, not only the matter energy tensor in \ref{conformal} becomes singular but also the entropy density takes negative value. So, these kind of processes has to be ruled out from the theory. This is ensured by the second condition  $Rf'(R)-f(R) \geq 0$~\cite{Jacobson:1995uq, Jacobson:1994qe}.
For $f(R)=R+\alpha R^2$ theory, this condition translates to be $\alpha>0$, which ensures the stability of the theory.  We would like to emphasize that, the above two inequalities hold in the physical frame and can also be realized without doing a conformal transformation.
\\

Further, by a change of variable\cite{Jacobson:1994qe} $\varphi = \beta ln f'(\phi)$, with $\beta = \sqrt{3/16\pi G}$, \ref{I2} becomes,

$$I_3= \int d^4 x \sqrt{-\bar{g}}[\frac{1}{16\pi G} \bar{R}- \frac{1}{2} \bar{\nabla}_\mu\varphi \bar{\nabla}_\nu\varphi - V(\varphi)+ e^{-2\beta^{-1}\varphi}L_m(\psi,e^{-2\beta^{-1}\varphi}\bar{g})]\label{I3}$$

Where $V(\varphi) = \frac{1}{16\pi G}e^{e^{-2\beta^{-1}\varphi}}(Rf'(\phi)-f(\phi))$. In the absence of matter the last term of $I_3$ vanishes and the action turns out to be that of a Einstein-scalar theory with some exotic potential $V(\varphi)$ minimally coupled to the curvature. In such case, for asymptomatically flat black hole solution the positivity of the potential i.e $V(\varphi)\geq 0$ or $Rf'(\phi)-f(\phi)\geq 0$ leads to a no-hair theorem\cite{Canate:2015dda}. However, when the above condition is violated, one needs to check for numerical solution for the existence of scalar hair. For the specific case of asymptotically flat spherically symmetric black hole solutions in various $f(R)$ models of interest, numerical evidences exists\cite{Canate:2015dda}, which shows the solutions to be non-hairy. But such result assumes spherical symmetry and can't be generalized unless the topology is known a priori. 
Inclusion of matter however, breaks the minimal coupling in \ref{I3} and in general one may not have a no-hair theorem. This is not surprising, since even in GR, there exists black hole solutions with scalar hair\cite{Delgado:2016jxq,Herdeiro:2014goa,Bekenstein:1998aw}, which doesn't obey any energy conditions. In such cases the topology theorem may not hold. However, in the spirit of Hawking's topology theorem in general relativity, we assume the matter-energy tensor to obey Dominant energy condition.
\\

Hence, under the above two physically reasonable conditions, motivated from the positivity of entropy density, no-hair theorems and stability of the theory, the Euler characteristics takes the form,

\begin{align}
\chi(\mathcal{H}) &= \int_\mathcal{B} \left[2\left(p_\mu p^\mu - \frac{d\theta}{ds}\right) + \frac{R f'(R)-f(R)}{f'(R)}  + \left(\frac{16 \pi T_{\mu \nu} k^\mu n^\nu}{f'(R)}\right)\right]  dA\nonumber \\
& +\int_\mathcal{B} \left[ \frac{2 k^\mu n^\nu \triangledown_\mu \triangledown_\nu f'(R) + 2\square f'(R)}{f'(R)} \right]dA \label{eu_ch1}
\end{align}

Where the positive definite terms are collectively written in the first integral. 
Now using the specific form of the induced metric $\beta_{\mu\nu}$, the second integral of above equation can be simplified  as,

$$2 k^\mu n^\nu \triangledown_\mu \triangledown_\nu f'(R) + 2\square f'(R) =\left(k^\mu n^\nu+k^\nu n^\mu +2 g^{\mu\nu}\right)\triangledown_\mu \triangledown_\nu f'(R) = \beta^{\mu\nu}\triangledown_\mu \triangledown_\nu f'(R)+ \square f'(R)$$

Extending the calculation further, one can easily show that,

\begin{equation}
\int_\mathcal{B} \frac{\beta^{\mu\nu}\triangledown_\mu \triangledown_\nu f'(R)}{f'(R)} dA= \int_\mathcal{B} \frac{\left[\triangledown^\mu f'(R)\right]\left[\triangledown_\mu f'(R)\right]}{f'(R)^2} dA \label{positive}
\end{equation}

Since, $k^\mu$ is a Killing field along the horizon, i.e, $k^\mu\triangledown_\mu f'(R)|_\mathcal{H} = 0$, the integrand on the right hand side of the above equation can be shown to be strictly positive. And, finally the Euler characteristics becomes,

\begin{align}
\chi(\mathcal{H}) = &\int_\mathcal{B} \left[2\left(p_\mu p^\mu-\frac{d\theta}{ds}\right) + \frac{R f'(R)-f(R)}{f'(R)} + \frac{\triangledown^\mu f'(R)\triangledown_\mu f'(R)}{f'(R)^2} + \left(\frac{16 \pi T_{\mu \nu} k^\mu n^\nu }{f'(R)}\right) \right]  dA \nonumber\\
&+\int_\mathcal{B} \left[ \frac{\square f'(R)}{f'(R)}\right]dA \label{eu_ch2}
\end{align}

\ref{eu_ch2} represents the final form of Euler characteristics in the physical frame. Terms in the first integral are separately positive definite and hence, integrated over a compact surface $\mathcal{B}$ gives a positive contribution to $\chi$. But, in general the sign of $\square f'$ is not certain. However, under the assumption, $\square f' \geq 0$, $\chi$ becomes positive and the topology turns out to be either toroidal or spherical. If the case of toroidal topology can be ruled out as in case of GR, then one left with a horizon cross-section in the $f(R)$ theory with spherical topology.\\ 

We should emphasize that $\square f' \geq 0$ is a sufficient condition which only leads to the fact that the Euler characteristics is positive semi-definite and therefore the topology of the horizon cross-section is either spherical or toroidal. We still can not rule out the toroidal topology using this condition. For this, we need either a generalization of the agreements of stability in GR or the work of\cite{Jacobson:1994hs}. Such a generalization may require the proof of topology censorship theorem\cite{Friedman:1993ty} for $f(R)$ gravity. Also, we need this condition to be valid only at the location of the horizon.\\

As emphasized earlier in this section, the result so far is in the physical frame. As we know, the $f(R)$ gravity can be transformed into Einstein gravity with an effective energy-momentum tensor by a conformal transformation. Such a procedure is useful to understand certain aspects of f(R) gravity. For example, in \cite{Jacobson:1995uq} , the conformal transformation technique is used to derive a classical second law for black holes in f(R) gravity. We would like to understand the emergence of the sufficient condition $\square f'(R) \geq 0$ also in the conformal frame which is related to the physical frame via a transformation given in \ref{con_tr}. Under such transformation, with $f'(R) = \omega^2$, some of the terms of interest transform as,

\[ \left\{
\begin{array}{ll}
     k^\mu\rightarrow \bar{k}^\mu = \frac{1}{\omega^2}k^\mu \\
n^\mu\rightarrow \bar{n}^\mu = n^\mu, \beta_{\mu\nu}\rightarrow \bar{\beta}_{\mu\nu}=\omega^2 \beta_{\mu\nu}\\

R\rightarrow \bar{R} = \frac{1}{\omega^2}R - \frac{6}{\omega^3} g^{\mu\nu}\nabla_\mu\nabla_\nu \omega \\

R(k,n)\rightarrow\bar{R}(\bar{k},\bar{n}) = \frac{1}{\omega^2}R(k,n)-\frac{1}{\omega^3}[2k^\mu n^\nu \nabla_\mu\nabla_\nu \omega - \square\omega] -\frac{1}{\omega^4}(\nabla_\mu\omega)(\nabla^\mu\omega)
\\
R(k,n,k,n)\rightarrow\bar{R}(\bar{k},\bar{n},\bar{k},\bar{n}) = \frac{1}{\omega^2}R(k,n,k,n)-\frac{2}{\omega^3} (k^\mu n^\nu)\nabla_\mu\nabla_\nu \omega + \frac{1}{\omega^4}(\nabla_\mu\omega) (\nabla^\mu\omega)
\\ \theta\rightarrow\bar{\theta} = \frac{1}{\omega^2}\theta +\frac{1}{\omega^4} k^\mu\nabla_\mu f'(R)
\end{array} 
\right. \]
\\

These transformations preserve the causal structure of space-time. Also, the Euler characteristics and hence the topology of the horizon cross-section remains invariant under such transformation. This implies, positivity of $\bar{\chi}$ in conformal frame guarantees the positivity of $\chi$ in physical frame. Therefore, one should expect to reproduce exactly the same condition for positivity of $\bar{\chi}$ as well.
\\

Since, $k^\mu$ is a Killing field only on the horizon, $k^\mu\nabla_\mu f'(R)|_\mathcal{H}$ identically vanishes and therefore $\theta$ and the shear $\sigma_{ab}$ remain zero under conformal transformation from a stationary space-time. As a result, a stationary Killing horizon retains its stationary Killing nature even in the conformal frame. However, since the Killing property of $k^\mu$ is specific to the horizon only, the rate of change of expansion along the auxiliary null normal $n^\mu$ changes under conformal transformation non-homogeneously as,

\begin{align}
\frac{d\bar{\theta}}{ds}= \frac{1}{\omega^2} \frac{d\theta}{ds} + \frac{1}{\omega^4}n^\alpha\nabla_\alpha(k^\mu\nabla_\mu f'(R))\label{inhom}
\end{align}
It is important to note that the second term in the right hand side need not be zero even on a Stationary Killing horizon as it refers to the derivative away from the horizon.
Also, we have
$$\int\left[\bar{p}^\mu \bar{p}_\mu -\frac{d\bar{\theta}}{ds}\right] d\bar{A} \longrightarrow \int \left[p^\mu p_\mu - \frac{d\theta}{ds} + \frac{\square\omega}{\omega} \right]dA$$

The above transformation holds upto a total divergence term of the form $d^\dagger p$. This can also be directly obtained from the conformal transformation of \ref{first_term}.\\

In conformal frame, the Euler characteristics then becomes,

\begin{equation}
\bar{\chi}(\bar{\mathcal{B}}) = \int_{\bar{\mathcal{B}}} 2\left[(\bar{p}|\bar{p})-\frac{d\bar{\theta}}{ds}\right] d\bar{A} + \int_{\bar{\mathcal{B}}} \bar{\mathcal{T}}(\bar{k},\bar{n}) d\bar{A}
\end{equation}\\

Here $\bar{\mathcal{T}}(\bar{k},\bar{n})$ represents the $(\bar{k},\bar{n})$ component of the effective energy-momentum tensor that appears in the $f(R)$ equation and it involves higher order curvature terms. The scalar product $\bar{p}^\mu \bar{p}_\mu$ is written as  $(\bar{p}|\bar{p})$.
In the conformal frame the field equation turns into Einstein's equation, with an effective energy-momentum tensor satisfying dominant energy condition which requires $f'(R) > 0$ and $R f'(R) - f(R) > 0$. Hence, $\bar{\chi}$ will turn out to be positive, provided that, $\bar{p}^\mu$ is a space-like vector and the event horizon in the conformal frame is the limit to the existence of outer trapped surfaces, i.e $\frac{d\bar{\theta}}{ds}|_{\bar{\mathcal{B}}}<0$. These conditions therefore guarantee the positivity of the Euler characteristics in the conformal frame which in turns lead to a positive Euler characteristics in the physical frame without any further condition!\\

This seems to be contradicting w.r.t the requirement of the sufficient condition derived by the direct calculation in the physical frame. To resolve this apparent contradiction, we first note that the quantities $\frac{d\bar{\theta}}{ds}$ and $(\bar{p}|\bar{p})$ do not transform homogeneously. This amounts to say that the condition for the non-existence of trapped surfaces outside the horizon is not a conformal invariant statement. As a result, although the Euler characteristics is positive in the physical frame with conditions: $f'(R) > 0$ and $R f'(R) - f(R) > 0$, this does not guarantee $\frac{d\theta}{ds} < 0$, and therefore does not exclude the existence of trapped surfaces in the domain of outer communication of the horizon. In fact, by substituting various conformal transformations, it is straightforward to show that the Euler characteristics in the conformal frame is,

\begin{align}
\bar{\chi}(\bar{\mathcal{B}}) &= \int_{\bar{\mathcal{B}}}2\left[  (p|p)-\frac{d\theta}{ds} +\frac{1}{2}\frac{\square f'(\phi)}{f'(\phi)} \right] dA \nonumber\\&+\int_{\bar{\mathcal{B}}} \left[ \frac{16\pi}{f'(\phi)}T(k,n) + \left(\frac{f''(\phi)}{f'(\phi)} \right)^2 (\nabla_a\phi) (\nabla^a\phi) + \frac{\phi f'(\phi)-f(\phi)}{f(\phi)}\right]dA \label{eu_ch final}
\end{align}

Since $\phi=R$, $\bar{\chi}$ in the conformal frame as given in the above equation is identical to $\chi$ in \ref{eu_ch2}. As a result, this expression is exactly same as the Euler characteristics in the physical frame, as one would expect. Therefore, the same sufficient condition $\square f'(R) \geq 0$ is needed if we impose no trapped surface condition in the physical frame.

\section*{Generalization to Higher Dimensions}
Now we want to extend the above result to higher dimensional black holes. In higher dimensions, the Euler characteristics of the horizon cross-section is no longer given by the expression  \ref{Eulerch} and the topological classification of the higher dimensional manifolds are more involved. Therefore, following ~\cite{Galloway:2005mf,Galloway:2011np}, we look for the Yamabe invariant instead of Euler characteristics, which is defined as,

\begin{equation}
\mathcal{Y}[\sigma] = \sup_{[\gamma]} \inf_{\bar{\gamma}\in [\gamma]} \frac{\int_\sigma S_{\bar{\gamma}}d\bar{\Sigma}}{\int_\sigma d\bar{\Sigma}}
\end{equation}

Here, $\sigma$ is a compact co-dimension two surface in a $(d+1)$ dimensional manifold $(\mathcal{M}^{d+1},g)$ with induced metric $\gamma$. $[\gamma]$ represents the conformal class of metrics of $\gamma$, i.e the set of all possible metrics related to $\gamma$ by a conformal transformation. Result of ~\cite{Galloway:2005mf} shows that, if $\sigma^{d-1}$ is a marginally outer trapped surface (which is the case for horizon), with $g$ being a solution of Einstein's equation and matter obeying dominant energy condition, then $\sigma$ admits positive scalar curvature and hence a positive Yamabe invariant. We follow the same procedure but for $f(R)$ gravity and use the corresponding field equation. For the sake of making the article self contained, we first briefly review the steps originally given by Galloway and Schoen \cite{Galloway:2005mf} and then use $f(R)$ equation to obtain our result. We show that the same sufficient condition is enough to guarantee the positivity of the Yamabe invariant of the horizon cross-section.
\\

We start by considering $\sigma$, to be a marginally outer trapped surface, i.e, expansion $\theta$ to be vanishing on $\sigma$ and positive outside. For our case, $\sigma$ will serve as the horizon cross-section. Being a co-dimension $2$- surface, it posses two normal directions and let $\nu$ and $u$ be the space-like and time-like normal to $\sigma$. The idea is to deform the surface $\sigma$ outward along $\nu$ in $V^d$, where $V^d$ is the $d$ dimensional space-like Cauchy surface which extends from the horizon to the space-like infinity. This is done with an initial deformation velocity $v=\phi \nu$ to get a surface $\sigma_t$ at some later time $t$. Here $\phi$ is some smooth function in $\sigma$. Then the null expansion varies as~\cite{Galloway:2005mf,Galloway:2011np, Cai:2001su},

\begin{equation}
\frac{\partial\theta}{\partial t}=\square\phi + 2\left<X,\nabla\phi\right> + \left( \frac{1}{2}S-G(u,k)-\frac{1}{2}|\chi_k|^2 + div X -|X|^2 \right)\phi
\end{equation}
Taking $Q = \frac{1}{2}S-G(u,k)-\frac{1}{2}|\chi_k|^2$, the above equation becomes,

\begin{equation}
\frac{\partial\theta}{\partial t}=\square\phi + 2\left<X,\nabla\phi\right> + \left( Q + div X -|X|^2 \right)\phi\label{variation}
\end{equation}

Here the Laplacian, gradient and divergences are taken on $\sigma$ with respect to the induced metric $\gamma$. $\chi_{ab}$ denotes the extrinsic curvature of $\sigma$ with respect to the outgoing null vector $k=u+\nu$ and   $S$ is the scalar curvature associated with $\sigma$. $X=-n_\alpha \nabla_A k^\alpha$ is a space-like vector on $\sigma$, with $'A'$ representing the spatial indices.  $G(u,k)$ is the $(u,k)$  component of Einstein tensor.

\ref{variation} can also be expressed as an eigenvalue equation,

\begin{equation}
\frac{\partial\theta}{\partial t}=L(\phi)=\lambda \phi
\end{equation}

Where $L$ is known as the stability operator. In general, $L$ is not self adjoint due to the divergence term it contains, but as shown in~\cite{6}, the principal eigenvalue of $L$ is always real and the principal eigenfunction $\phi$, associated with it is positive definite. Since $\sigma$ is a marginally outer trapped surface, $\frac{\partial\theta}{\partial t} >0$, i.e, $L(\phi)>0$. Following this inequality, one can show,

\begin{equation}
\int_\sigma |\nabla\phi|^2 + Q\phi^2\geq 0, \forall \phi\in C^\infty(\sigma)
\end{equation}

Now consider the operator $L_1 = -\square + Q$ and the eigenvalue equation $$L_1\phi = \lambda_1\phi$$ 

The principal eigenvalue of $L_1$ is given by,

\begin{equation}
\lambda_1(p) = \inf \frac{\int_\sigma \phi L_1(\phi) d\Sigma}{\int_\sigma \phi^2 d\Sigma}
= \inf\frac{\int_\sigma |\nabla\phi|^2 + Q\phi^2 d\Sigma}{\int_\sigma \phi^2 d\Sigma} \label{principal ev}
\end{equation}

From \ref{principal ev} it is clear that the principal eigenvalue of the operator $L_1$ is always positive and the principal eigenfunction, let $\phi_p$, can always be chosen to be strictly positive.

In this setting, the scalar curvature $\bar{S}$ of surface $\sigma$, with respect to a metric $\bar{\gamma}= \phi_p^{2/d-2}\gamma$ is of the form~\cite{Galloway:2011np};
\begin{align}
\bar{S}\nonumber &= \phi_p^{-2/d-2}\left[(-2\square+2Q)\phi_p + 2G(u,k)\phi_p +|\chi_k|^2\phi_p  +\frac{d-1}{d-2}\frac{|\nabla\phi_p|^2}{\phi_p^2} \right]\\
&=\phi_p^{-2/d-2}\left[\lambda_1(p)\phi_p + 2G(u,k)\phi_p +|\chi_k|^2\phi_p  +\frac{d-1}{d-2}\frac{|\nabla\phi_p|^2}{\phi_p} \right]\label{scalarc1}
\end{align} 

All other terms in \ref{scalarc1} except $G(u,k)$ is positive definite. In case of GR, we would substitute $T(u,k)$ in the place of $G(u,k)$. But, for $f(R)$ gravity, we use the corresponding field equation and obtain,
\begin{equation}\label{G}
G(k+n,k) = \frac{8\pi T(u,k) + k^\mu k^\nu \nabla_\mu\nabla_\nu f'(R) + \left(Rf'(R)-f(R)\right)+\square f'(R) + k^\mu n^\nu \nabla_\mu\nabla_\nu f'(R)}{f'(R)}
\end{equation} 
Now, on the horizon we find that, 

$$k^\mu k^\nu \nabla_\mu\nabla_\nu f'(R) = k^\mu\nabla_\mu(k^\nu\nabla_\nu f')-(k^\mu\nabla_\mu k^\nu)\nabla_\nu f' = 0$$

Also, using the same result as in the case of $ 3 + 1$ dimensions, the last two terms on the right hand side of \ref{G} can be further simplified as,

$$\frac{\square f'(R)+k^\mu n^\nu \nabla_\mu \nabla_\nu f'(R)}{f'(R)} = \frac{1}{2}\frac{\square f'(R)}{f'(R)}+ \frac{\beta^{\mu\nu}\nabla_\mu \nabla_\nu f'(R)}{f'(R)}$$

When integrated over a compact surface the last term gives a positive contribution to the Yamabe invariant(see \ref{positive}).
So using the same argument presented in previous section, we can conclude the scalar curvature of $\sigma$ over a conformal class of metric as given in \ref{scalarc1} is strictly positive, provided $\square f'(R)\geq 0$, which yields a positive Yamabe invariant.\\

Manifolds with positive scalar curvature and hence positive Yamabe invariant has been extensively studied by numerous authors\cite{2017arXiv170700902H,Hallam:2017xkt,2015arXiv150303803A}. Although there is not much rigid restriction on the topology of higher dimensional manifolds, in five dimension the allowed topology is limited. For instance as discussed in\cite{Galloway:2011np}, the compact 3-dimensional cross-section admitting positive Yamabe invariant  of $4+1$ dimensional black hole is diffeomorphic to $S^3$ or $S^1\times S^2$ or connected sum of these. As a result, the differential condition $\square f'(R) \geq 0$ will restrict the topology of a five dimensional black hole in f(R) gravity to be either $S^3$ or $S^1\times S^2$ or connected sum of these.

\section*{Discussion}

The admissible topology of black hole horizons could be a probe to distinguish gravity theories from general relativity. The possible topology of asymptotically flat black holes in general relativity is severely constrained by the Hawking's topology theorem. This may not be the case for theories with higher curvature corrections. Such theories may admit asymptotically flat black hole solutions with different topologies. The kinematics of such black holes could be completely different from the case of general relativity. This provides an interesting approach to look for the physics beyond the Einstein gravity. For example, the quasi normal modes of a stationary asymptotically flat black hole could be sensitive to the topology of the horizon and in principle may be used to detect the violation of general relativity.\\

This motivates to generalize the Hawking's topology theorem for simplest higher curvature theories, namely the $f(R)$ gravity. We found that the generalization requires a sufficient condition $\square f'(R) \ge 0$ on the horizon along with the usual dominant energy condition. This differential condition is required to be valid only at the horizon. To understand more about this condition, let us consider the case: $f (R) = R +\alpha \, R^p$. Then, the sufficient condition reduces to $\square (R^{p-1}) \ge 0$. 
For $ p = 2$, we can understand such a constraint further in terms of a condition of the trace of the energy-momentum tensor. This can be realized by taking the trace of the field equation for $R+\,\alpha R^p$ gravity to obtain,
\begin{equation}
\square(R^{p-1}) = \frac{8\pi T + R + \alpha(2-p)R^p}{3\alpha p}\label{boxfR}
\end{equation}

Where $T$ denotes the trace of energy-momentum tensor.
Let us concentrate on the special case of $p=2$, i.e for $f(R)= R+\, \alpha R^2$ theory. This limiting case also has been particularly studied as a model of inflation in the literature\cite{Starobinsky:1979ty,Kehagias:2013mya,Asaka:2015vza,Artymowski:2015mva}.
In this limit the sufficient condition reduces to $\square R \geq 0$. Therefore, any black hole solution of $R +\alpha \, R^2$ gravity for which the curvature is an harmonic function of space-time will have either spherical or toroidal topology in $3+1$ dimensions and horizon cross-section with positive Yamabe invariant in higher dimensions. 
For this particular case, \ref{boxfR} takes the form,

\begin{equation}
\square R = \frac{R+8\pi T}{6\alpha}\label{boxR}
\end{equation}

Further, we note that, for the case of $R+\alpha R^2$ theory, $f'(R) = 1 + 2\alpha R >0$ and hence the Ricci scalar is bounded from below, i.e $R> -1/2\alpha$. Using this bound we conclude that, \ref{boxR} yields a non negative value of $\square R$, provided a sufficient condition $T\geq (1/ 16 \pi \alpha)$ holds. Therefore, for the positivity of Euler characteristics, the requirement of $\square R \geq 0$ (which is a condition on the curvature) is now reduced to a condition on matter, given in terms of the coupling constant. Hence, this condition along with the dominant energy condition determines the topology of horizon. In $3 + 1$ dimension, if the energy-momentum tensor obeys dominant energy condition and the trace is bounded from below by $1/ 16 \pi \alpha$, the topology is either $S^2$ or $S^1 \times S^1$. In higher dimension, the same condition makes the Yamabe invariant positive. So, for $ R + \alpha \, R^2$ gravity, the dominant energy condition along with a trace energy condition $ T \geq 1/ 16 \pi \alpha$  is required for the validity of Hawking's topology theorem.\\

However, as one can see from \ref{boxfR}, such a simplification doesn't happen for $p>2$. Hence, the topology of horizon may not be straightforwardly obtained from the matter sector alone and the additional condition of $\square f'(R)\geq 0$ is required.\\

Another important issue with our derivation is the validity of the rigidity theorem in $f(R)$ gravity. We have assumed that the event horizon in the stationary space-time is also a Killing horizon. This is a consequence of Hawking's rigidity theorem\cite{1,2,Hollands:2006rj} which is only true for general relativity and there is no generalization of this result in higher curvature gravity including the $f(R)$ theories.  But, at least in $3+1$ dimensions, this is a reasonable assumption as the original derivation of Hawking in $3+1$ dimensions may be extended to $f(R)$ gravity by conformal transformations. \\

Finally, note that even with the differential condition, we can not rule out the Toroidal topology and a detailed study involving the generalization of the topology censorship theorems is needed to settle this issue.

\section*{Acknowledgement}
We thank Avirup Ghosh for extensive discussion and comments. We also thank Anjan Sen for discussion. Research of SS is supported by the Department of Science and Technology, Government of India under the SERB MATRICS Grant (MTR/2017/000399). MR thanks INSPIRE-DST, Government of India for a Junior Research Fellowship.

\end{document}